\documentclass[10pt,journal,compsoc]{IEEEtran}

\usepackage{lmodern}
\usepackage[utf8]{inputenc}
\usepackage[T1]{fontenc}
\usepackage[nocompress]{cite}
\usepackage[pdftex]{graphicx}
\usepackage{url}
\usepackage{listings}
\usepackage{booktabs}
\usepackage{tabularx}
\usepackage{algpseudocode}
\usepackage{framed}

\graphicspath{{figures/}}
\DeclareGraphicsExtensions{.pdf,.png}

\algnewcommand{\LeftComment}[1]{\Statex $//$ \textit{#1}}
\algnewcommand{\Variable}[2]{\Statex \textbf{Variable} #1: #2}
\algnewcommand{\Def}[2]{\State \textbf{def} #1: \textit{#2}}

\newcommand{\code}[1]{\textsf{\small{}#1}}
\newcommand{\doc}[1]{\textsf{\footnotesize``#1''}}
\newcommand{\para}[2]{\vspace{2mm}\noindent\textbf{#1: }#2}

\begin{document}
\lstset
{
   language=Java,
   basicstyle=\scriptsize\sffamily,
   showstringspaces=false,
   frame=none,
   morekeywords={enum, assert},
   tabsize=2
}

\title{Generating Unit Tests for Documentation}

\author{Mathieu Nassif, Alexa Hernandez, Ashvitha Sridharan, and Martin P. Robillard
\IEEEcompsocitemizethanks{\IEEEcompsocthanksitem The authors are with the School of Computer Science, McGill University, Montr\'{e}al, Canada.%
\protect\\
E-mail: \{mnassif, martin\}@cs.mcgill.ca,\protect\\
E-mail: \{alexa.hernandez, ashvitha.sridharan\}@mail.mcgill.ca}%
\thanks{Manuscript received ...; revised ...}}

\markboth{Submitted to IEEE Transactions on Software Engineering}%
{Nassif et al.: Generating Unit Tests for Documentation}

\IEEEtitleabstractindextext{%
\begin{abstract}
Software projects capture information in various kinds of artifacts, including source code, tests, and documentation. Such artifacts routinely encode information that is redundant, i.e., when a specification encoded in the source code is also separately tested and documented. Without supporting technology, such redundancy easily leads to inconsistencies and a degradation of documentation quality. We designed a tool-supported technique, called DScribe, that leverages redundancy between tests and documentation to generate consistent and checkable documentation and unit tests based on a single source of information. DScribe generates unit tests and documentation fragments based on a novel template and artifact generation technology. By pairing tests and documentation generation, DScribe provides a mechanism to automatically detect and replace outdated documentation. Our evaluation of the Apache Commons IO library revealed that of 835 specifications about exception handling, 85\% of them were not tested or correctly documented, and DScribe could be used to automatically generate 97\% of the tests and documentation.
\end{abstract}

}

\maketitle

\IEEEdisplaynontitleabstractindextext
\IEEEpeerreviewmaketitle

\IEEEraisesectionheading{\section{Introduction}\label{s:intro}}

\IEEEPARstart{M}{ature} software frameworks and libraries are usually complemented by extensive test suites and reference documentation. For example, the Apache Commons Math project release 3.6 is supported by 4467 tests and 215\,176 words of method reference documentation. Although unit tests and reference documentation serve different purposes, their creation involves expressing the same or similar information in different software artifacts, which must then be kept consistent. An example of a pervasive case is that of a function that throws a specific type of exception when supplied with an invalid argument. Normally, such behavior should be described in the function's documentation, and tested by a unit test. Ideally, the test and the documentation would be consistent.

As this simple scenario illustrates, current practices for testing and documenting reusable software assets exhibit three inter-related problems. First, manually-created tests and documentation are often \textit{redundant}. In turn, this redundancy introduces the risk of \textit{inconsistencies} between a documented specification and the exercise of the corresponding behavior in a test. Finally, in situations where many functions in a library exhibit similar constraints (e.g., on input validation), the redundancy between tests and documentation exacerbates the \textit{repetitiveness} of the testing and documentation effort.

The goal of our research is to leverage the redundancy and repetitiveness of information in software artifacts to reduce the amount of developer effort, as well as the threat of inconsistencies. To advance towards this goal, we investigate a solution that explores a new synergy between template-based unit test and documentation generation.

Although, at an abstract level, generating unit tests may seem relatively straightforward, realizing this idea in practice required addressing many new technical challenges with original solutions. We explored this design space by fully developing a prototype technique, called DScribe, that can generate unit tests for Java systems.

DScribe is a tool-supported technique for transforming facts about methods between different types of equivalent representations. DScribe relies on a database of fact templates, and users invoke a template to instantiate a specific fact
about a method into a unit test and corresponding block of documentation. To realize this functionality, the design of DScribe incorporates, among others, a new template definition language and original algorithms for aggregating related fragments of documentation into a cohesive unit.

As a research project focused on engineering design, our assessment of DScribe focused on gaining an understanding of the potential usefulness of the approach, its applicability, and its limitations. A study revealed that 85\% of the specifications about exceptions thrown by the methods of the Apache Commons IO library are either untested, undocumented, or both. In addition, the investigation revealed that DScribe could have prevented 97\% of these inconsistencies. In a wider study of the applicability of DScribe, we found that 42\% of the tests in three additional Apache commons projects captured at least one unit of specification, which means that a significant amount of tests need to be kept consistent with documentation.

The main contribution of this paper is the complete design and implementation of a prototype technique for generating unit tests for documentation. Although this technique only represents one point in a wide design space, we also contribute numerous insights about the rationale for important design and implementation solutions that can inform future work in that direction. We also contribute three empirical studies that provide different insights on the general potential for generating unit tests for documentation. Although they leverage our work on DScribe, the studies are not specific to the tool, and thus the observations they generated can provide insights that go much beyond the application of a given prototype.

This article is organized as follows. In Section~\ref{s:overview} we provide a general overview of DScribe, followed by sections that supply the details on the two key aspects of the approach: templates (Section~\ref{s:templ_invoc}) and generative technology (Section~\ref{s:testdoc}). Section~\ref{s:evaluationOverview} is an overview of the empirical assessment of the work, followed by the details of a usefulness study (Section~\ref{s:UsefulnessStudy}), a validation study (Section~\ref{s:ValidationStudy}), and a qualitative empirical study of the limitations of the approach (Section~\ref{s:LimitationsStudy}). Section~\ref{s:related} presents the related work and Section~\ref{s:conclusion} concludes the paper.

\subsubsection*{Research Artifacts}

Our contributions are complemented by an on-line appendix, which contains the source code of DScribe, details of the evidence collected as part of the studies, as well as additional details on the implementation of the technique.

\vspace{2mm}
\noindent
\url{https://github.com/prmr/DScribe-Research}

\section{DScribe Overview}
\label{s:overview}

\begin{figure}
	\centering
	\includegraphics[width=0.7\linewidth]{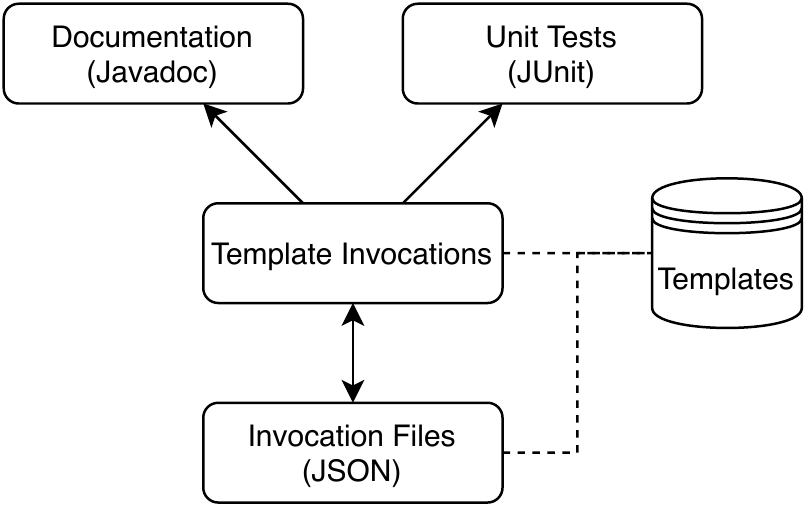}
	\caption{Information representations in DScribe. Rounded rectangles show different representations of facts, and arrows indicate transformations supported by DScribe. Dashed lines indicates a dependency to a set of templates.}
	\label{f:overview}
\end{figure}

At its core, DScribe is an approach to transform \textit{facts} about software elements between different types of equivalent representations. In the current implementation of DScribe, facts must relate to a single public Java method, called the \textit{focal method}.

For example, a fact could be that if the focal method receives the value \code{null} as argument, it throws a \code{Null\-Pointer\-Exception}. The method declaration encodes this fact by providing an implementation. However, the fact can also be redundantly encoded in other representations, such as unit tests or natural language documentation.

In addition to their redundant representations, some facts can have a repetitive structure when minor variations of the same fact apply to multiple methods. DScribe comprises a database of fact \textit{templates}, where each template captures one common kind of facts. For example, one template captures facts with the structure ``method $X$ throws an exception of type $Y$ when the argument is $Z$.'' A template includes \textit{placeholders} that provide the flexibility to use the template in different contexts.

Users \textit{invoke} a template to instantiate a specific fact about a method. A \textit{template invocation} provides the values for the placeholders of the template for a given focal method. Template invocations serve as a common basis for all representations of facts. Figure~\ref{f:overview} summarizes the representations supported by DScribe and the relations between them. In practice, template invocations are employed to produce two commonly-used representations of facts: header comments used for \textit{documentation} (Javadoc) and \textit{unit tests} using the JUnit framework.

This synergistic combination of automatically generated tests and documentation mitigates the respective weaknesses of both representations of information: The documentation is made checkable and traceable to source code (via its connection to unit tests), and the latent documentation captured by unit tests is made explicit and easily accessible (as documentation). A further benefit of DScribe's approach is that, once generated, the tests and documentation are well-formed artifacts fully independent from the generation framework.

\begin{table}
	\centering
	\caption{Technical and implementation challenges involved in the development of DScribe. The last column indicates the section of this article that discusses the challenge.}
	\label{t:challenges}
	\begin{tabular}{@{}llr@{}}
		\toprule
		\textbf{Component}       & \textbf{Challenge}                    & \textbf{Sect.}       \\
		\midrule
		\multicolumn{3}{c}{\textit{Technical}} \\
		Templates                 & Capturing kinds of facts              & \ref{s:templates}   \\
		Invocations               & Capturing minimal information         & \ref{s:invocations} \\
		Invoc. $\to$ Tests        & Ensuring compilability                & \ref{s:ptypes}      \\
		Invoc. $\to$ Doc.         & Reducing clutter                      & \ref{s:aggregation} \\
		\midrule
		\multicolumn{3}{c}{\textit{Implementation}} \\
		Templates                 & Serializing template information      & \ref{s:templates}   \\
		JSON                      & Designing a lossless readable format  & \ref{s:invocations} \\
		Invoc. $\to$ Tests        & Proper code style and                 & \ref{s:test-integ}  \\
		~                         & \quad integration with existing tests &                     \\
		Invoc. $\to$ Doc.         & Traceability and                      & \ref{s:doc-integ}   \\
		~                         & \quad integration with source code    &                     \\
		\bottomrule
	\end{tabular}
\end{table}

Our research into the development of DScribe required solving a number of design challenges, but also experimenting with alternative solutions to implementation challenges. Table~\ref{t:challenges} summarizes these challenges.

Defining the structure of templates and invocations were the two first challenges. The guiding principle behind their design was to facilitate the generation of tests and documentation while avoiding unnecessary or redundant information. Because these two components were novel aspects of DScribe, they also involved significant implementation challenges.

Invocations and their JSON representations are roughly equivalent, so the bidirectional transition between them is straightforward. Given a template invocation, the generation of unit tests is, for the most part, an implementation rather than a design challenge. Nevertheless, ensuring that the generated unit tests are compilable is not trivial, and requires the definition of types for placeholders. Integrating the generated tests and documentation with other artifacts of the system is also challenging, especially with the constraint to minimize repetitiveness of the generated documentation. The solution to this latter challenge led to the design of a novel intermediate representation for documentation.

Because implementation challenges are less relevant to the research, we only mention them briefly for completeness. Our publicly available implementation of DScribe provides a fully developed prototype solution to these implementation challenges.

A key design principle of DScribe was to avoid any possibility for imprecise inference. Past research has proposed various inference techniques to extract and generate information, often using one representation to generate another~\cite{RBK2012a}. These techniques are useful to \textit{discover} information initially unavailable to developers. However, in the context of our research goal, their limitation is that they require developers to validate the outcome of the inference process. In contrast, DScribe's aim is to effectively leverage information that has already been specified by developers precisely and unambiguously.

\section{Templates and Invocations}
\label{s:templ_invoc}

Templates and their invocations are the main innovations of DScribe. They are a new form of documentation for software systems better suited to represent common facts without repetitiveness and unnecessary information. To achieve this objective, templates encapsulate as much as possible the information that would otherwise be repeated between methods and their associated artifacts (documentation and test), and template invocations encapsulate the remaining information that is method-specific.

\subsection{Template Definition Language}
\label{s:templates}

\begin{figure}
	\begin{lstlisting}
/** $method$ throws an exception of type $ex$
 *  when $state$.
 */
@Template("Example")
@Types($ex$=EXCEPTION, $state$=EXPR, $factory$=METHOD)
@Test
public void test$method$_$state$() {
	$class$ instance = $factory$();
	try {
		instance.$method$();
		fail();
	} catch ($ex$ e) {}
}
	\end{lstlisting}
	\vspace*{-1em}
	\caption{Example of a template.}
	\label{f:template}
\end{figure}

DScribe's templates are, literally, templates for tests and documentation: a DScribe template contains a partial abstract syntax tree (AST) of a unit test and a partial natural language description. These two elements are only partial because they contain placeholders, each identified by a unique name. Thus, a template is exactly the aggregation of a list of placeholders, an AST rooted at a method declaration, and a natural language description.

Figure~\ref{f:template} shows an example of a template. The Java code defines the template AST and the header comment is the template description, with placeholders identified by a surrounding pair of dollar signs (\$). The template expresses the fact that a method throws an exception when its implicit argument is in a specific state. The specific method, exception type, and state are all placeholders of the template, as well as the declaring type of the method and the name of a factory method.

The value of a template (i.e., the information that it captures) is more than the sum of its parts (partial AST and description). A template explicitly associates two representations of a specific kind of facts: how to describe the fact in documentation, and how to test it. It is this association that allows DScribe to transform different representations of the same information, without relying on inference techniques.

For a template to be effective, its AST and description must be self-contained, but also flexible. A developer should be able to understand the purpose of the template by reading the description and the AST, but also to apply this template to various contexts, or methods. A catalog of templates can thus serve not only to generate tests and documentation, but also as a knowledge base for the development community.

An important design principle for templates was to avoid any reliance on a prescribed coding style, except for the requirement that each unit test focuses on a single test case about a single method. For example, with our design, templates can enfore any convention for unit test names.

\subsubsection*{Implementation Decisions}
\label{s:templatedb}

Templates are collected in a catalog that consists of a set of parsable Java files. Each method declared in these files and identified with the  \code{@Template} annotation corresponds to a template, as in Figure~\ref{f:template}. The AST of the method declaration becomes the AST of the template, and the header comment of the method becomes its description. Each placeholder is a legal Java identifier that begins and ends with a dollar sign (\$).\footnote{Although it is a legal character for identifiers, the Java Language Specification discourage the use of dollar signs in usual code~\cite[\S3.8]{GJS2020}, thus reducing the probability of collisions between templates and actual code.} The  \code{@Types} annotation declares the list of placeholders,\footnote{It also assigns a type to each placeholder. Placeholder types are designed mostly for unit test generation, so we discuss them in Section~\ref{s:ptypes}.} except for a few predefined placeholders that refer to the properties of the focal method, such as \code{\$method\$} (its name) and \code{\$class\$} (its declaring type). Template authors attribute the template's name as the only argument of the \code{@Template} annotation (e.g., \code{Example} in Figure~\ref{f:template}).

The motivation for expressing templates using legal Java code was for the template format to be familiar to Java developers. This makes templates more readable, and consequently the knowledge they capture more accessible. The format also allows template authors to leverage their usual tools to create and edit templates. Finally, this format facilitates the creation of new templates from existing tests and documentation: a developer only needs to clone the existing test and documentation, and replace specific values with placeholders.

\subsection{Template Invocations}
\label{s:invocations}

A template invocation records the application of a template to a focal method.
Invocations require an \textit{invocation context} that consists of the signature of the focal method and values for placeholders. The method signature is required to correctly link the generated assets (tests and documentation). It also provides the values of the few predefined placeholders (e.g., \code{\$method\$}), and a default package and Java type from which other placeholder values can be resolved.

The remainder of the invocation context is the set of values to assign to the template's placeholders. Following the principle that the generation of tests and documentation should be as transparent for the user as possible, DScribe replaces placeholders with the user-provided values with as little transformation as possible to make the test compile or the documentation sensible. Thus, the values supplied to the template invocation are not expressions to be evaluated by the generation engine, but expressions to be substituted verbatim for the placeholders.

\subsubsection*{Implementation Decisions}

We made the arbitrary decision to format invocation files using JSON. This format allows users to directly read, write, and edit invocation files with any text editor. Users can also use many existing tools to manipulate JSON objects more effectively. Finally, many JSON libraries support the implementation of a straightforward serialization and deserialization of invocations.

To keep invocation files as concise as possible, they contain only the necessary information. Hence, in contrast to template files, which are mostly self-contained, invocation files are not. They depend on the definition of the relevant templates to generate useful information. Future versions of DScribe could include tools to make invocation files easier to create, read, and edit, but this implementation challenge is left for future work.

\section{Unit Test and Documentation Generation}
\label{s:testdoc}

\begin{figure}
	\includegraphics[width=\linewidth]{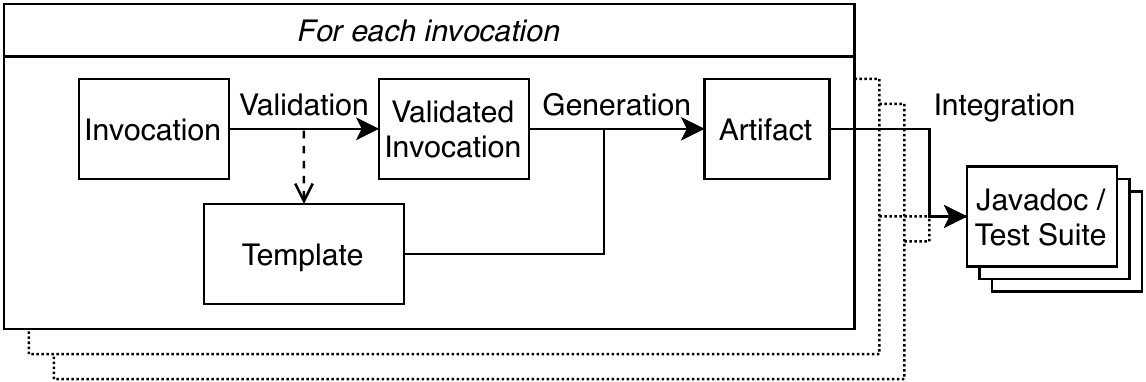}
	\caption{High-level overview of tests and documentation generation}
	\label{f:gen-overview}
\end{figure}

The ultimate goal of DScribe is to manage external fact representations: unit tests and Javadoc documentation. The generation of both kings of artifacts follows a similar three-step process summarized in Figure~\ref{f:gen-overview}.

For each individual invocation, the first step is to identify the template being invoked and validate the values of the invocation. The validation step ensures that the invocation refers to a valid template, that its focal method exists, and that each placeholder value can be correctly substituted. Validating placeholder values is especially important, and challenging, for the generation of compilable unit tests. This challenge motivated the design of a small type system for placeholders that can guide developers in properly invoking templates (see Section~\ref{s:ptypes}).

If the invocation passes the validation step, the second step is to generate the artifact (i.e., unit tests or documentation fragment) by substituting each placeholder with the value that the invocation provides.

After generating all artifacts individually, the third and final step is to combine them and integrate them with the rest of the system. Generated unit tests are integrated within the existing test suite, and generated documentation is integrated directly in the source code as header comments for documentation. In addition to inserting the generated artifacts into the project without disrupting the rest of the system, the integration step is also responsible for removing outdated generated artifacts, and aggregating similar generated artifacts to reduce repetitiveness. This aggregation is especially important for documentation, because unnecessary clutter will have a negative impact on readability. To solve this aggregation challenge, we designed a novel, easily interpretable structured format to express facts, which also allows for a trivial yet effective aggregation of similar facts.

\subsection{Placeholder Types}
\label{s:ptypes}

During the generation of tests, the value that each placeholder replaces is subject to different syntactic rules, depending on the location in which the placeholder appears in the template. For example, in the assignment \code{Object x = \$p1\$(\$p2\$)}, the first placeholder, \code{\$p1\$}, can never be replaced by an integer literal. Placeholder types can help avoid such errors. Each different type defines a specific set of rules that apply to a placeholder based on its location in the template.

\begin{figure}
	\begin{lstlisting}
$type$ x = $expr$;
Object y = x.$method$($exprlist$);
System.out.println(y.$field$);
throw new $exception$();
	\end{lstlisting}
	\vspace*{-1em}
	\caption{Example of a partial AST template with placeholders of different types}
	\label{f:ex-types}
\end{figure}

The placeholder types DScribe supports are \code{TYPE}, \code{EXCEPTION}, \code{METHOD}, \code{FIELD}, \code{EXPR}, and \code{EXPR\_LIST}. Figure~\ref{f:ex-types} shows usage examples of a placeholder of each type.

Placeholder values of type \code{TYPE} and \code{EXCEPTION} must be the qualified name of an existing Java type in the build path of the system. Additionally, \code{EXCEPTION} placeholders must inherit from the \code{Throwable} class. A qualified name is necessary to resolve the Java type, but also to insert it in the template without an associated \code{import} statement. However, to reduce unnecessary effort, if the Java type is declared in the same package as that of the focal method, only the simple name is required. Placeholders of type \code{METHOD} and \code{FIELD} replace a method or field name, respectively. Placeholder values of type \code{EXPR} must be syntactically legal Java expressions. Similarly, the \code{EXPR\_LIST} type can be used for placeholders that replace a variable number of expressions, usually for the arguments of a method invocation in the template (see the placeholder \code{\$exprlist\$} in Figure~\ref{f:ex-types}).

For the types \code{METHOD}, \code{FIELD}, \code{EXPR}, and \code{EXPR\_LIST}, DScribe does not resolve the identifiers used in the placeholder values, and so it does not verify that the method or field exists, or that the expressions refer to existing variables. Therefore, it is possible that DScribe will generate unit tests with compilation errors due to unresolved symbols or incompatible types (e.g., if the value of \code{\$expr\$} is incompatible with the Java type \code{\$type\$} in Figure~\ref{f:ex-types}). This limitation is a necessary condition to allow developers to create templates reusable in various contexts, and it is mitigated by the fact that the compiler of the test suite will detect these errors.

The context of the invocation provides a few predefined placeholders, including \code{\$method\$}, \code{\$class\$}, and \code{\$package\$} for the focal method (or constructor) and its declaring type and package, respectively. These placeholders do not require an explicit value from the user, as the value is derived from the context. Therefore, they do not have a placeholder type.

\subsection{Integration of Unit Tests with Existing Test Suite}
\label{s:test-integ}

Integrating the generated unit tests with an existing test suite is mostly an implementation challenge. Each template defines, using placeholders, the name of the class and package where to place the generated test. Thus, DScribe groups together all tests that go in the same file, and writes one file for each such test class. Because each unit test is independent, and because a large number of similar tests are not an issue for test suites (they are meant to be read by the testing framework), there is no further aggregation to perform.

To avoid corrupting the manually written testing code, but also remove outdated tests generated by previous executions, DScribe places all generated tests in a separate folder, defined by the user. Therefore, manually and automatically generated test code can coexist in separate folders, and template invocations can leverage scaffolding, such as stub objects and helper methods, from the main test suite. When a user executes DScribe again, old test classes (in DScribe's folder) are simply overwritten by the new ones, thus removing any outdated test.

\subsection{Information Aggregation}
\label{s:aggregation}

Some methods can have multiple similar facts that need to be tested and documented. For example, Java's \code{Math.log(double)} method returns the natural logarithm of its argument. However, the natural logarithm, as a real-valued function, is only mathematically defined for positive numbers, so the method needs to define special behaviors for other values. In particular, \code{Math.log} returns \code{NaN} for negative numbers and \code{NaN} itself. Thus, a DScribe user would create two separate invocations for these two special cases, and DScribe would then generate two similar documentation fragments, e.g., \doc{If \code{a} is \code{NaN}, the result is \code{NaN}} and \doc{If \code{a} is negative, the result is \code{NaN}}.

Because DScribe inserts the documentation fragments directly into the header comment (Javadoc) of the focal method, such repetitive fragments are undesirable. Instead, it is preferable to use a single sentence to express both cases, as currently found in the documentation: \doc{If the argument is NaN or less than zero, then the result is NaN}~\cite{JavadocMath}.

\begin{figure}
	\centering
	\includegraphics[width=\linewidth]{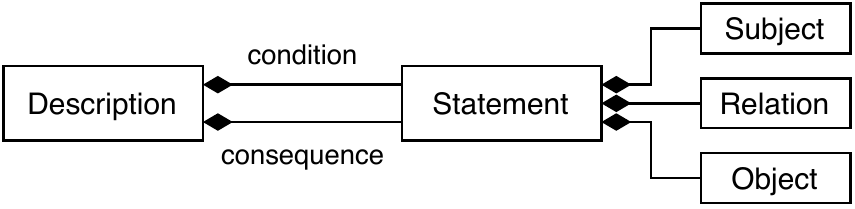}
	\vspace*{-2em}
	\caption{UML description of the structure of documentation fragments}
	\label{f:info-structure}
\end{figure}

To support the aggregation of similar fragments, we designed a structure to express facts about method behavior. Figure~\ref{f:info-structure} summarizes this structure. Each documentation fragment is divided into two statements, a \textit{condition} and a \textit{consequence}, with the interpretation that if the condition is true, then the consequence must also be true. For example, for \code{Math.log}, the condition \doc{\code{a} is negative} is associated with the consequence \doc{the result is \code{NaN}}. Furthermore, each statement is itself composed of three parts, a \textit{subject}, a \textit{relation}, and an \textit{object}, similarly to Resource Description Framework (RDF) triples~\cite{Lassila1999}, with the interpretation that the relation applies from the subject to the object. Thus, the condition of the previous example would have \doc{\code{a}} as subject, \doc{is} as relation, and \doc{negative} as object.

Altogether, this structure requires template authors to divide the natural language description associated with the template into six parts (i.e., the subject, relation, and object for both the condition and consequence) which map naturally to most specifications. Each part can be any text (with placeholders), to allow as much flexibility as possible. Using this structure, DScribe can aggregate similar fragments without the need for natural language processing techniques. If two or more fragments share the same condition (resp. consequence), DScribe aggregates the consequences (resp. condition) of those fragments. When aggregating statements, DScribe can also avoid the repetition of a common subject, relation, and/or object.

In our example, DScribe would initially generate two fragments, \doc{If \code{a} is \code{NaN}, then the result is \code{NaN}} and \doc{If \code{a} is negative, then the result is \code{NaN}}. These two fragments have the same consequence (\doc{the result is \code{NaN}}), so DScribe will aggregate the conditions. Both conditions share the same subject (\doc{\code{a}}) and relation (\doc{\code{is}}), so the aggregated condition will become \doc{\code{a} is \code{NaN} or negative}. Thus, both fragments become the single fragment \doc{If \code{a} is \code{NaN} or negative, then the result is \code{NaN}}.

Although the six-part structure naturally maps to most method behaviors, to account for cases that are impossible to express with this structure, either or both statements can be replaced with free form text, or the complete fragment can be free-formed. Using these special provisions, however, limits the ability of DScribe to aggregate similar fragments.

\subsection{Integration of Documentation with Code}
\label{s:doc-integ}

DScribe inserts the generated documentation fragments into the Javadoc header comment of the focal method, creating the header comment if necessary. Thus, the documentation is explicitly linked to the code element (i.e., method) it applies to. This is made possible by the explicit link between a template (which captures the documented information) and its focal method.

DScribe uses the custom \code{@dscribe} Javadoc tag to mark generated fragments from those previously written by developers. Each (aggregated) fragment is prefixed by its own tag to avoid a single, large paragraph that is hard to read.

The custom \code{@dscribe} tags not only clearly indicate to the user which statements are automatically generated, and backed by unit tests, it also allows DScribe to keep the documentation up-to-date: when the user modifies the template invocations, DScribe can remove all previous \code{@dscribe} tags and regenerate them with the new information. This process does not impact the manually written documentation at all, unless users manually modify the content of generated tags.

\section{Overview of the Empirical Assessment}
\label{s:evaluationOverview}

\begin{table*}
	\centering
	\caption{Overview of the empirical assessment of unit test generation for documentation}
	\label{t:studies}
	\begin{tabular}{@{}lclllr@{}}
		\toprule
		\textbf{Study} & \textbf{Sect.} & \textbf{Subject} & \textbf{Scope} & \textbf{Purpose} & \textbf{RQs} \\
		\midrule
		Usefulness & \ref{s:UsefulnessStudy} & Commons IO (root package) & Exceptions & Assess DScribe's ability to prevent inconsistencies & 1, 2 \\
		Validation & \ref{s:ValidationStudy} & Commons Math, Lang, Config. & Tested specifications & Validate the results of the usefulness study & 1 \\
		Limitations & \ref{s:LimitationsStudy} & 5 open source projects & All & Understand DScribe's limitations & 2 \\
		\bottomrule
	\end{tabular}
\end{table*}

Leveraging our implementation of DScribe, we conducted a multi-pronged empirical investigation of key aspects relating to the generation of unit tests for documentation.
The investigation sought to answer the following research questio\textsl{}ns.
\begin{enumerate}
	\item[RQ1] To what extent is information in source code, unit tests, and documentation inconsistent?
	\item[RQ2] To what extend can we leverage DScribe templates to automatically test and document behaviors of focal methods?
\end{enumerate}

In this investigation, we followed a three-stage process, with each stage constituting a cohesive study of its own:

\begin{enumerate}
	\item \textbf{Usefulness Study:} We conducted an \textit{in-depth} study of the usefulness of DScribe in a \textit{narrow context} (Section~\ref{s:UsefulnessStudy});
	\item \textbf{Validation Study:} We conducted a \textit{multi-case} study to validate the findings of the first study in a \textit{broader context}  (Section~\ref{s:ValidationStudy});
	\item \textbf{Limitations Study:} It is customary to discuss the limitations of proposed software engineering techniques. In our case we also conducted an empirical study to better understand the limitations of DScribe in diverse scenarios (Section~\ref{s:LimitationsStudy}).
\end{enumerate}

Table~\ref{t:studies} provides an overview of the empirical work described in this article. For all three studies, the complete and detailed results are publicly available in our on-line appendix.

\section{Usefulness Study}
\label{s:UsefulnessStudy}

We investigated the usefulness of DScribe to prevent inconsistencies. The objective of the investigation was twofold. First, it aimed at understanding the nature of the problem of information repetitiveness and redundancy. Second, it aimed at assessing the potential of DScribe to avoid future inconsistencies by automating the generation of repetitive and redundant information.

\subsection{Usefulness Study Design}
\label{ss:usefulnessStudyDesign}

To study information inconsistency and redundancy, it was necessary to define what constitues a cohesive unit of information about a method. The information relevant to methods typically includes units of specification regarding, among others, exceptions, parameter types, edge cases, and return types. We chose as our unit of analysis such a \textit{unit of specification}.

In the general case it requires a significant amout of manual effort to isolate and fully understand even just a few units of specifications in unfamiliar code. To make this case study tractable, we narrowed the scope to a particular type of unit of specification that is well-defined: units of specification about exceptions, which are relevant in almost all systems. For a single method, its source code, associated unit tests, and documentation should present the same information about thrown exceptions. Thus, an \textit{exception specification unit} (ESU) is inconsistent if there is any divergence in its expression in its associated artifacts (code, documentation, or tests). This definition includes the cases where an artifact omits the ESU.

As the subject of the case study, we chose the Apache Commons IO library (version 2.6, commit 11f0abe). This library consists of utility functions and classes, each mostly independent of the others, with well-defined ESUs. It is also extensively documented and tested. Because the library contains a total of 152 top-level Java types, an amount which precludes an in-depth analysis of each method, we focused only on the public types in the root package \code{org.apache.commons.io}. We also excluded deprecated, abstract, and exception types, which resulted in eleven remaining classes and a total of 293 public, non-deprecated methods. Table~\ref{tab:scope} presents an overview of these classes, including the number of ESUs and templates identified for each class.

\begin{table}
	\begin{center}
		\caption{Number of methods, exception specification units (ESUs), and instantiated DScribe templates per class under investigation}
		\label{tab:scope}
		\begin{tabular}{@{}lrrr@{}}
			\toprule
			\textbf{Class} & \textbf{Methods} & \textbf{ESUs} & \textbf{Instances} \\
			\midrule
			\textbf{ByteOrderMark} & 8 &  6 & 6\\
			\textbf{ByteOrderParser} & 1 & 1 & 1\\
			\textbf{Charsets} & 4 & 2 & 2\\
			\textbf{EndianUtils} & 30 & 67 & 67\\
			\textbf{FileCleaningTracker} & 7 & 8 & 8\\
			\textbf{FileDeleteStrategy} & 5 & 4 & 4\\
			\textbf{FileUtils} & 95 & 403 & 386\\
			\textbf{FilenameUtils} & 33 & 30 & 28\\
			\textbf{HexDump} & 2 & 5 & 5\\
			\textbf{IOUtils} & 101 & 315 & 295\\
			\textbf{LineIterator} & 7 & 8 & 8\\
			\midrule
			\textbf{Total} & 293 & 849 & 810\\
			\bottomrule
		\end{tabular}
	\end{center}
\end{table}

For each method declared in the classes under study, barring deprecated and private ones, one of the authors identified all ESUs present in at least one of the documentation, test suite, and source code. The identified ESUs include not only exceptions directly thrown by the method under investigation, but also those thrown by nested calls, which explains the large effort involved in eliciting the ESUs. For each ESU, the investigator noted the type of exception thrown, the state that triggers the exception, which of the source code, test suite, and/or documentation captured the ESU, and which DScribe template could be used to generate a unit test and documentation for this exception, creating the template if necessary. For the latter, if no template could capture the ESU, the investigator recorded the reason instead. Of the 849 ESUs identified, the investigator was not able to verify the correctness of 14 with respect to the source code. These 14 cases are included in Table~\ref{tab:scope}, but we omitted them from the rest of the study.

\subsection{Results and Discussion}

\begin{table}
	\begin{center}
		\caption{Presence of exception specification units (ESUs) in documentation and unit tests. For each value, the number after the ``+'' sign indicates the number of ESUs that are not present in the source code.}
		\label{tab:UOSE}
		\begin{tabular}{@{}lrrr@{}}
			\toprule
			& \textbf{In Doc.} & \textbf{Not in Doc.} & \textbf{Total} \\
			\midrule
			\textbf{In Test}     & 122+1            &  29+0                & 152\\
			\textbf{Not in Test} & 458+9            & 216+0                & 683\\
			\midrule
			\textbf{Total}       & 590              & 245                  & 835\\
			\bottomrule
		\end{tabular}
	\end{center}
\end{table}

To answer RQ1, Table~\ref{tab:UOSE} summarizes the degree to which identified ESUs are consistent across the artifacts of Commons IO, by comparing the number of ESUs described in the documentation (In Doc.) and tested by the test suite (In Test). The results highlight the pervasiveness of information inconsistencies in Commons IO: 85\% of the identified ESUs are missing in at least one of the documentation, test suite, or source code. An even more concerning observation is that the overwhelming majority (82\%) of ESUs are untested, which increases the risk of documentation becoming silently inaccurate. This risk of documentation becoming silently inaccurate is already exemplified from the 10 cases where ESUs in documentation are not traceable to the source code. In this case, use of DScribe would also remediate the 19\% of tested ESUs that are absent from the documentation, presumably by accident.

In some cases, an ESU was only partially or vaguely described in the documentation. Of the 590 ESUs present in documentation, 22 (4\%) did not include the type of exception thrown, and 115 (19\%) only described the input state that triggers the exception in broad terms, or aggregated multiple invalid states. A recurring example of such broad documentation in the \code{FileUtils} class is \doc{IOException - if source [file] is invalid}. Here, an API user is left wondering about the various specific invalid input states that may trigger the exception, such as a file that does not exist or that is a directory. Such cases, which decrease the usefulness of documentation, would be avoided by DScribe.

\begin{table}
	\begin{center}
		\caption{Number of times that each DScribe template was instantiated}
		\setlength{\tabcolsep}{0.7\tabcolsep}
		\label{tab:templates}
		\begin{tabular}{@{}lrlr@{}}
			\toprule
			\multicolumn{2}{c}{\textbf{Invocations}} & \multicolumn{2}{c}{\textbf{Not Invoked}} \\
			Template           & Count               & Reason          & Count \\
			\midrule
			Static             & 237         (28\%)  & Inaccurate Doc. & 10 (1\%)  \\
			NotStatic          &   2 \enspace (0\%)  & Unable to Test  & 15 (2\%)\\
			MessageStatic      & 547         (66\%)  & ~               & ~ \\
			MessageNotStatic   &  19 \enspace (2\%)  & ~               & ~ \\
			MessageConstructor &   5 \enspace (1\%)  & ~               & ~ \\
			\midrule
			\textbf{Total}     & 810         (97\%)  & \textbf{Total}  & 25 (3\%)\\
			\bottomrule
		\end{tabular}
	\end{center}
\end{table}

To answer RQ2, the investigator created the necessary templates and invocations to capture as many ESUs as possible. Table~\ref{tab:templates} shows the resulting templates, and the number of invocations for each of them, as well as the reason why we could not invoke any template for some ESUs. The fact that 97\% of the identified ESUs could be captured by a template invocation confirms DScribe's potential to avoid future information inconsistencies. Each such invocation would lead to a unit test and a documentation fragment. Failing unit tests would instantly flag invocations inconsistent with the source code, thereby alleviating the burden of having to maintain ESUs in multiple artifacts manually.

The results also show that DScribe's ESU templates are highly reusable. Almost all ESUs (94\%) were supported by only two templates, \code{Static} and \code{MessageStatic}. Thus, the overall relative cost of template creation is low. In our case, 810 ESUs (97\%) were instantiated using only five templates. The five templates vary depending on the different types of focal methods (static, non-static, and constructor), and whether to verify the message of the exception. The templates \code{NotStatic} and \code{MessageNotStatic}, designated for non-static focal methods, were used less often as most methods under investigation were static. Similarly, the \code{MessageConstructor} template was not widely used because few ESUs were identified for constructors.

In addition to these results, we observed the use of three alternative patterns to test exceptions. Namely, using a \code{try-catch} block with JUnit's \code{fail} method, using JUnit's \code{assertThrows} method, and using helper methods to verify the type and message of an exception. It is thus evident that developers leverage recurrent templates naturally, but inconsistently. This inconsistency hinders readability and, consequently, maintainability. DScribe helps standardize the consistent use of recurrent templates, thus enhancing the quality of test suites.

We were not able to instantiate ESUs in only 25 cases (3\%), due to two main reasons. The first one was the presence of inaccurate ESUs in the documentation, i.e., statements in the documentation that did not reflect the actual behavior of the method. While it is possible to instantiate these ESUs, it would lead to failing unit tests and outdated documentation. We did not instantiate the other ESUs because we could not produce input states that would trigger the target exception. For example, it is not possible to ensure that an \code{InputStream} instantiated inside a method, rather than passed as an input parameter, produces an \code{IOException} when it is read. The majority of these cases were also not tested in the test suite.

\subsection{Threats to Validity}
\label{s:threats-usefulness}

Two of the authors performed all annotations. It is possible that the investigators may have missed some ESUs, or misinterpreted the purpose of a test or behavior of a method, as they are not part of the development team for the library under test. We mitigated this threat by selecting a library that requires little specialized knowledge. Additionally, the methods of utility libraries are usually self-contained and can be understood without knowledge of the system as a whole. Nevertheless, the annotations may still reflect the investigators' experience. To ensure verifiability, we include the complete results of our study in our on-line appendix.

A threat to external validity stems from our decision to focus on exception handling. We do not expect that this context would generalize to all types of units of specification. Moreover, we only investigate eleven classes from Commons IO. The results may not generalize to the library as a whole, let alone other systems. Similarly, our results are dependent on our selection of templates. Different templates may not be as reusable. Nevertheless, the study demonstrates the usefulness of DScribe in at least one realistic software development context, as we applied it to the popular Commons IO library, from which we can analytically generate to similar software components.

\section{Validation Study}
\label{s:ValidationStudy}

The results presented in Section~\ref{s:UsefulnessStudy} clearly indicate that information inconsistency across source code, documentation, and unit tests is a clear issue for exception handling in the Commons IO project. To refine the answer to RQ1, we performed a multi-case study to validate and expand the findings of the initial case study.

\subsection{Validation Study Design}

Because identifying all units of specifications from the \textit{source code} of a method is both effort-intensive and subjective (due to the ambiguity of what constitutes a single ``unit''), this second study focused on the units of specifications found in \textit{unit tests}. This design restricts the scope of the validation study to testable (and tested) specifications, but it is necessary to make the findings reliable.

As the subjects of the validation study, we selected the three Apache Commons projects with the most unit tests: Math (version 3.5, commit d7d4e4d, 3757 tests), Lang (version 3.8.1, commit 2ebc17b, 3086 tests), and Configuration (version 2.4, commit 61732d3, 2554 tests). We chose to again study Apache Commons projects for the same reasons outlined in Section~\ref{s:UsefulnessStudy}. We randomly sampled tests uniformly from the total population of 9397 tests. For each sampled test, one author manually identified the focal unit of the test, using the test's name, the \textit{Last Call Before Assert}~\cite{VanRompaey2009}, and comments. Because we were focusing on specifications about methods in the production code, we rejected tests whose focal unit was not a single method (e.g., multiple methods, or a class or field), or if the focus was ambiguous. We also rejected degenerate cases (e.g., empty, deprecated, or auto-generated tests). We continued the sampling until we gathered a set of 370 viable tests, rejecting a total of 93 unsuitable tests. This sample size is sufficient to support a generalization of proportions of tests computed on the sample to the whole population within a 5\% confidence interval at the 0.95 level.

For each test in the sample, one author noted if the test captured at least one unit of specification about the focal method (some complex tests actually tested multiple inputs), writing it down to ensure it was well-defined. Each specification was expressed as \textit{If X, then calling the method will do Y}, to avoid considering all information (including, e.g., usage examples) as a specification. For each identified unit of specification, the investigator then noted whether it was described in the documentation, and if so, if the description was only partial and broad, or complete and explicit.

\subsection{Results and Discussion}

\begin{table}
	\begin{center}
		\caption{Number of unit tests capturing at least one specification (documented or not). Percentages are computed with respect to each project.}
		\label{tab:results}
		\begin{tabular}{@{}lrrrrr@{}}
			\toprule
			\textbf{Project} & \multicolumn{3}{c}{\textbf{Information Present}} & \textbf{No Info.} & \textbf{Total} \\
			~ & Doc. & Partial & Not doc. & ~ & ~ \\
			\midrule
			\textbf{Config.} & 9 \enspace (9\%) & 3 (3\%) & 16 (15\%) & 76 (73\%) & 104 \\
			\textbf{Lang} & 55 (39\%) & 10 (7\%) & 20 (14\%) & 57 (40\%) & 142 \\
			\textbf{Math} & 17 (14\%) & 3 (2\%) & 24 (19\%) & 80 (65\%) & 124 \\
			\midrule
			\textbf{Total} & 81 (22\%) & 16 (4\%) & 60 (16\%) & 213 (58\%) & 370 \\
			\bottomrule
		\end{tabular}
	\end{center}
\end{table}

Table~\ref{tab:results} presents, for each project, the number of tests that captured self-contained information about a specification of its focal method (\textit{Information Present}), or not (\textit{No Info.}), and whether the information was completely included in the documentation (\textit{Doc.}), partially or broadly (\textit{Partial}), or not mentioned at all (\textit{Not doc.}).

Overall, 42\% of tests captured at least one unit of specification, which means that a significant amount of tests need to be kept consistent with documentation. This proportion is even higher (60\%) for Lang. For Configuration and Math, undocumented specifications amount to over half of the tested specifications, a situation that the use of DScribe prevents. In the case of Lang, although the lack of consistency is less significant, the use of DScribe would reduce the effort required to produce and maintain the more extensive documentation.

Multiple factors can explain the absence of unit of specification in the remaining 58\% of tests. In many cases, a test was simply verifying that under ``usual'' inputs, a method behave as it should. For example, the test \code{Kendalls\-Correlation\-Test\-.test\-Simple\-Reversed()} in Math simply validates that the correlation computed in a specific (normal) scenario is correct. Other cases, however, were more ambiguous: some tests captured at least a partial unit of specification, but the complete information was obscured by external references or ambiguous names. For example, \code{Test\-Data\-Configuration\-.test\-Get\-Byte\-Array()} in Configuration follows some recognizable patterns, but depends on values from configuration files referred to as \code{byte.list1}, \code{byte.list2}, etc. In such cases, the investigator used a conservative strategy and marked the test as capturing no specification. Nevertheless, refactoring the tests, or generating them with DScribe, could make them more self-contained, thus improving their quality. Numbers reported in Table~\ref{tab:results} should thus be regarded as lower bounds of the effective values.

The investigation of the sampled tests also revealed interesting use cases for DScribe outside the scope of this study. For example, the documentation of \code{Str\-Builder\-.as\-Tokenizer()}, from the Lang project, contains a usage example that is very similar to the test \code{Str\-Builder\-Test\-.test\-As\-Tokenizer()}. In such situation, developers could also use DScribe to generate usage examples for which the correctness is guaranteed by passing unit tests. This study also revealed the presence of incorrect documentation, such as that of the method \code{Dfp.reciprocal()}. Its unit tests, however, capture the correct behavior. By generating unit tests and documentation together, DScribe reduces the amount of brittle documentation that silently becomes inaccurate.

\subsection{Threats to Validity}

As for the usefulness study, an author performed all annotations, which leads to the same threats outlined in the Section~\ref{s:threats-usefulness}. However, as it is common in case studies, this procedure was necessary to obtain detailed insights that require a degree of interpretation. For verifiability, we include the complete results in our on-line appendix.

The target systems are a collection of mostly independent utility methods and classes, with extensive test suites, from the same organization as the usefulness study. We do not expect that this context would generalize to all systems. We are aware of this limitation, and we scope our claims accordingly. Nevertheless, the evaluation shows evidence of a considerable amount of information inconsistencies in a realistic and significant software development context.

\section{Limitations Study}
\label{s:LimitationsStudy}

The usefulness study showed evidence of the potential effectiveness of DScribe in one particular context, in which 97\% of the identified exception specification units could be captured by template invocations. However, to better answer RQ2 in a general context, we performed a qualitative multi-case study specifically to elicit the strengths and limitations of a template-based approach for generating unit tests and documentation.

\subsection{Limitations Study Design}

To ensure a variety of contexts, we selected five open source projects that are at least 15 years old and that vary in their development style, target audience, and application domain: Freemind (version 1.1.0, commit 643c55c), Eclipse Platform UI (version 4.9.0, commit d6d8a6a), Weka (version 3.9.3, commit r14866), Apache Tomcat (version 9.0.11, commit r183513), and Hibernate ORM (version 5.3.2, commit 35806c9).

One author annotated a subset of the test suite of each project. For each test, the investigator answered the question \textit{What are the technical factors that would enable or prevent the generation of similar unit tests from templates?} To help answer this question, the investigator noted the unit under test, purpose, format, and recurrent patterns for each test, in addition to the enabling and hindering factors.

\begin{table}
	\begin{center}
		\caption{Five open source subjects of the limitation study}
		\label{tab:systems}
		\begin{tabular}{lrrr}
			\toprule
			\textbf{System}	& \textbf{Prod. Files} & \textbf{Test Files} & \textbf{Inspected Files} \\
			\midrule
			Freemind  &  379 &   26 & 18 \\
			Eclipse	  & 3933 & 1669 & 49 \\
			Weka      & 1614 &  253 & 20 \\
			Tomcat    & 1402 &  475 & 16 \\
			Hibernate & 3845 & 5647 & 12 \\
			\bottomrule
		\end{tabular}
	\end{center}
\end{table}

To achieve maximal purposive sampling, instead of annotating a fixed subset of each project, the investigator iteratively included more unit tests to the sample until reaching saturation, which we defined as when three consecutive iterations generated no new noted observations. Each iteration consisted of selecting a package with at least three classes at random from the test suite of a project, then selecting three random classes (or more if the classes or package are small enough) from that package, and annotating all tests from these classes. The investigator analyzed one project at a time, moving to the next once saturation was reached for one. For Freemind, which only contains two test packages, the investigator annotated all unit tests from the root package of the test suite. Table~\ref{tab:systems} shows the number of production and tests Java files for each project, in the order they were annotated, as well as the number of inspected test files.

The investigator initially used an open coding process~\cite{Miles2013} to annotate each test. After completing the open coding, and after a preliminary analysis of the initial codes, the investigator systematically re-coded each test using a closed code catalog.

\subsection{Results and Discussion}

We identified eight technical factors that can impact the ability to generate unit tests from templates or the qualities of the generated tests. We discuss these factors at a high level in this section, but the interested reader can find multiple concrete examples of each factor in our on-line appendix. Although these factors outlined several limitations for using DScribe in different contexts, they also revealed simple strategies to work around these limitations, which can improve the quality of the generated tests. Furthermore, these limitations can provide insights about new features to add to DScribe in future work.

\para{Generic Variable Names} A template-based approach requires the use of recurrent, generic names for local variables in unit tests (e.g., \code{input} and \code{expected}), as opposed to names specific to the test context (e.g., \code{baseString}, \code{encodedString}). Although only a minority of the studied tests used such generic identifiers, we believe generic names can have a beneficial impact on the readability of the test suite, as it allows unfamiliar readers to understand new tests quickly by identifying recurrent important aspects. Thus, despite being a limitation of template-based approach, this factor can be beneficial in the long term.

\para{Structured Test Names} An important strength of a template-based approach is the ability to standardize and facilitate the use of conventions. In particular, although the name of unit tests does not impact its behavior, it is considered good practice to use meaningful names, usually following a fixed convention. As an extreme example of a highly-structured name, all test names in Tomcat's class \code{Check\-Out\-Thread\-Test} match the pattern \code{test\-(DBCP|Pool)\-Threads\-(10|20)\-Connections\-(10|20)\-(Validate)?\-(Fair)?}.

\para{Recurrent Complex Operations} A common limitation of template-based approaches is the diverging implementations of similar operations. For example, many tests verify that the content of a generated object matches that of the expected result, but the implementation of this verification depends on the structure of the objects. Thus, although the tests follow the same high-level patterns, they cannot be generated from the same template. However, we observed that some tests encapsulated such recurrent complex operations into helper methods with generic but meaningful names, such as \code{assert\-Content\-Equal}. The use of such helper methods can mitigate this limitation, and increase the readability of test suites. Nevertheless, relying excessively on helper methods can be detrimental. As an extreme, but not unique, example, Tomcat's helper method \code{Test\-EL\-Parser\-.do\-Test\-Parser} encapsulates all operations of multiple tests. This leads to a very complex logic that is harder to write and read than if the different tests were decoupled, and this helper method can only be used for testing a single class.

\para{Complex Assertions} Some tests require complex assertion structures. For example, testing methods that rely on inversion of control may need to nest assertions inside mock objects, and to call seemingly unrelated methods to trigger the assertions. Such structures can severely limit the applicability of templates, and thus the usefulness of a template-based approach. Thus, these complex cases remain mostly outside the scope of DScribe, or similar approaches. However, if the same pattern of complex assertion is often needed, a template-based approach can encourage developers to create the necessary scaffolding and have a more systematic approach to test complex behaviors.

\para{Testing Preconditions} Several tests include assertions to verify the input state of tested objects before the method under test is performed.\footnote{This practice is debated among developers, with some arguing that preconditions should be the object of separate tests. However, without taking a stance in this debate, given that at least some developers may want this feature, we consider its importance.} Although there is no reason a template could not include these early assertions, in many cases, the assertions are specific to the tested objects and relevant to only some test case, so they are not well suited for template-based generation. A simple workaround this limitation, however, is the use of factory methods to create the tested objects in the right input state, and move any necessary early assertion to these methods.

\para{Constrained Resources} Tests that rely on constrained resources, such as connections to external servers, multiple threads, or even read and write operations to the file system, may need to perform additional setup and cleanup operations to avoid corrupting the resources, as well as special precautions to control errors originating from the constrained resources themselves (e.g., trying a second time to connect to a server if the first time fails). These operations often create deviations from recurrent templates, thus multiplying the number of required patterns to account for each possible deviation. Different mitigation strategies can limit the negative impact of these operations, such as an efficient use of ``setup'' and ``teardown'' methods (using JUnit's \code{@Before} and \code{@After} annotations), but the right choice depend on the nature of the constrained resource.

\para{Different Units Under Test} In the sampled set of tests, the unit under test was not always a single method. Some tests focused on a whole class, whereas others focused on validating a single field. For example, Freemind's test \code{Html\-Conversions\-Test\-.test\-End\-Content\-Matcher} validates the expected behavior of a regular expression encoded in a constant field. Currently, DScribe assumes that the focus of each generated test is a method, so the generation of these other tests would be outside its scope. However, this is only a design decision for the prototype, and extensions of DScribe could include other types of units under test.

\para{Variety of Test Purpose} Tests in a test suite serve various purposes. Some simply test the usual behavior of a unit with specific examples, whereas others focus on exceptional behaviors or corner cases. Although it is important to test the former cases, it is likely that only the latter cases will need to be documented. Thus, a template-based approach should be able to generate documentation for only some templates to avoid clutter. More importantly, some tests verify the integration of various components in more complex scenarios (i.e., integration tests, which are not technically unit tests, but can still be found in the same test suites), and others are specifically tailored to a specific bug or case of regression. These kinds of tests are clearly outside the scope of template-based approaches, as they each require the execution of a specialized sequence of actions.

\subsection{Threats to Validity}
\label{s:threats-prestudy}

The case study relied on the identification of testing patterns, a subjective concept relative to the experience of every developer. Hence, the conclusions may reflect the personal experience of the investigator. This experimental design choice was necessary because the data analysis required a very high initial effort investment to study the systems, and a consistent point of view from one system to the other. Thus, the coding procedure could not be packaged into multiple sets of data to be labeled by independent coders. Furthermore, hypothetical external coders would have to be extensively trained to have the in-depth knowledge of the template-based approach required for the task, which would re-introduce the risk of bias. To mitigate this risk, the on-line appendix contains several concrete examples supporting each of our conclusions.

\section{Related Work}
\label{s:related}

The difficulty of maintaining high-quality documentation~\cite{Lethbridge2003, Fluri2007, Tan2007, Ratol2017} led to a vast exploration of \textbf{automated documentation generation} approaches. Techniques proposed in prior work involve static~\cite{Sridhara2011} and dynamic~\cite{Sulir2017} analysis of the body of methods, as well as their context~\cite{McBurney2014}, and different techniques are tuned to document either classes~\cite{Moreno2013}, methods~\cite{Sridhara2010, Nahla2015}, or method parameters~\cite{Sridhara2011}. Techniques also differ in the kind of documentation they generate, such as specifications~\cite{Ammons2002, Shoham2008, LeGoues2009}, program invariants~\cite{Ernst1999}, test summaries~\cite{Panichella2016, Zhang2016}, and usage scenarios and examples~\cite{Acharya2007, Buse2012}. However, a common limitation of such \textit{fully}-automated techniques is that the correctness and usefulness of the generated documentation is limited by the underlying heuristics and the information these heuristics rely on. Furthermore, some manual selection is ultimately required as the unconstrained generation of large amounts of information can end up diluting important insights within more trivial information. Being semi-automated, DScribe leverages developer effort, rather than replacing it, keeping developers in control of what information is added to the system.

A vast number of techniques have also been proposed to \textbf{automatically generate tests}. Notable early work includes CUTE~\cite{Sen2005} and DART~\cite{Godefroid2005}, which introduced the concept of \textit{concolic} testing. Concolic testing couples a symbolic and concrete execution of a program to explore the space of inputs that will trigger different responses from the program. Thummalapenta et al.~\cite{Thummalapenta2009} generate test cases by extracting sequence of method calls to create relevant input states. Pacheco et al.~\cite{Pacheco2007} proposed Randoop, a technique to generate test cases by randomly creating sequences of execution, with a feedback loop to inform the next generations. Fraser and Zeller~\cite{Fraser2012} follow a more systematic random generation approach by leveraging mutation operators, and using genetic algorithms to optimize the test suite. Taneja and Xie~\cite{Taneja2008} leverage the version history of a project to create test cases. Other techniques focus only on the generation of test cases that can crash a system~\cite{Csallner2004}, that apply to multi-threaded code~\cite{Nistor2012}, or that map to the system's UML diagrams~\cite{Offutt1999}. However, automated test generation techniques suffer from a similar problem as documentation generation techniques: In order to completely remove developers from the generation process, the techniques are susceptible to false positives, which in turn require human effort to filter out. In contrast, DScribe involves developers in the generation process so that no single person is required to sift through a large output after the generation.

The value proposition of DScribe, however, extends beyond the generation of tests and documentation. An important benefit is the \textbf{automated traceability links} of the generated artifacts to the method they complement. Documentation traceability is a challenging problem~\cite{Marcus2003, Antoniol2002}, but it is a prerequisite to validate the correctness of the documentation, another challenging problem~\cite{Zhou2017, BenCharrada2012}. DScribe offers a way to solve both problems by linking documentation not only to its focal method, but also to a unit test, such that if a change in the behavior of the method happens, the failing unit test will also flag the specific fragment of documentation as incorrect. This solution is similar to that of behavior-driven development (BDD)~\cite{Soeken2012}, a methodology derived from test-driven development~\cite{Beck2003}. BDD recognizes the documentation potential of testing code, and BDD frameworks such as JBehave~\cite{JBehave.org} offer a way to integrate documentation fragments directly into unit tests, so that documentation is again backed by passing tests. However, developers are still responsible for writing both the testing code and documentation fragments, a repetitive and redundant effort.

Finally, our research is related to that of \textbf{code pattern mining}, which parses large corpora of source code to identify regularities in the usage of various type of code elements (e.g., functions). The objective of these techniques is to identify specifications~\cite{Allamanis2014, Li2005, Acharya2007}, and in particular violations of these implicit specifications, or design patterns~\cite{Dong2009, Pandel2010}. Future work can leverage a similar approach to automatically generate DScribe templates, to further reduce the initial burden of developers.

\section{Conclusion}
\label{s:conclusion}

Motivated by the observation that documentation and testing code often capture redundant and repetitive information, we designed a technique, called DScribe, to allow developers to decouple aspects of unit testing and documentation that relate to repetitive specifications from the aspects specific to each instance. This technique can partially relieve developers of the burden of maintaining a consistent and extensive documentation and test suite, while also encouraging the use of collectively agreed upon templates to reduce unnecessary variability in these artifacts.

A three-phase investigation of the inconsistencies in selected mature software projects revealed their pervasiveness in testing code and method documentation, with 85\% of the specifications about exceptions thrown by the Apache Commons IO methods either untested, undocumented, or both. In addition, the investigation revealed that DScribe could have prevented 97\% of these inconsistencies in a favorable context. Finally, our empirical assessment of DScribe includes rich descriptions of the technical characteristics of software projects that facilitate or hinder the application of DScribe, thus providing insights on the potential costs and benefits of introducing the technique in different contexts.

\section*{Acknowledgments}

This work was funded by NSERC.

\bibliographystyle{IEEEtran}
\bibliography{IEEEabrv,bibliography}

\begin{IEEEbiography}[{\includegraphics[width=1in,height=1.25in,clip,keepaspectratio]{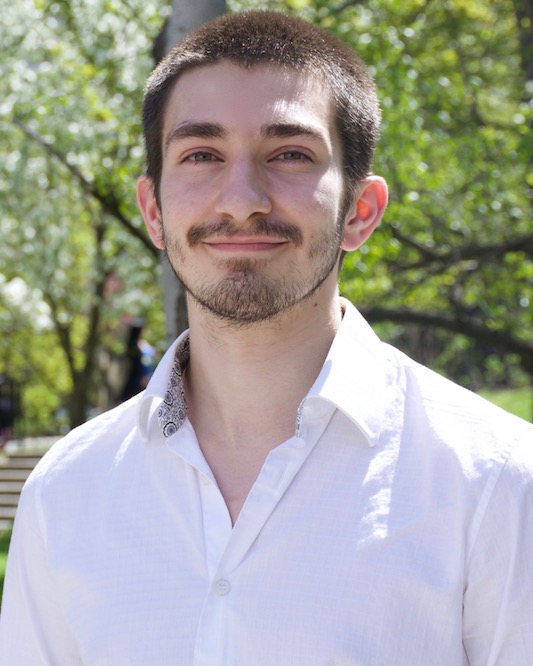}}]{Mathieu Nassif} is a Ph.D. student in Computer Science at McGill University, under the supervision of Martin Robillard. His research focuses on the extract, representation, and manipulation of knowledge in software systems to optimize the contribution of developers to the system. Mathieu received his M.Sc. in Computer Science from McGill University and his B.Sc. in Mathematics from Universit\'{e} de Montr\'{e}al.
\end{IEEEbiography}

\begin{IEEEbiography}[{\includegraphics[width=1in,height=1.25in,clip,keepaspectratio]{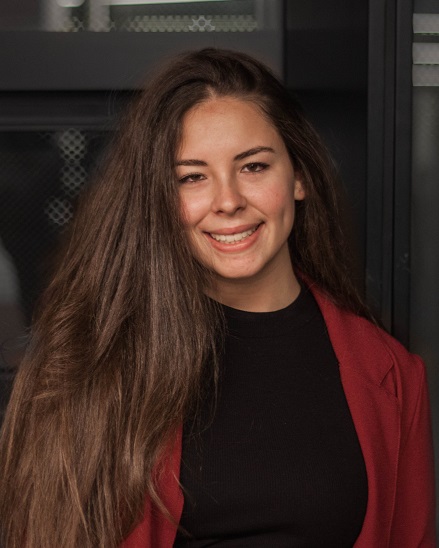}}]{Alexa Hernandez} is an incoming M.Sc. student in Computer Science at McGill University. Her research interests include software design, maintenance, and evolution. Alexa received a B.A. in Computer Science at McGill University, where she worked under the supervision of Martin P. Robillard.
\end{IEEEbiography}

\begin{IEEEbiography}[\vspace*{-2em}{\includegraphics[width=1in,height=1in,clip,keepaspectratio]{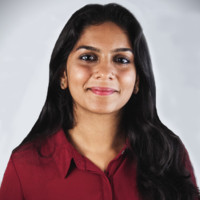}}]{Ashvitha Sridharan} is a software engineer optimizing the edge network at Shopify. Her research interests include software design, maintenance, and evolution. Sridharan received a B.Sc. Computer Science at McGill University, Montreal, where she worked under the supervision of Martin P. Robillard.
\end{IEEEbiography}

\begin{IEEEbiography}[{\includegraphics[width=1in,height=1.25in,clip,keepaspectratio]{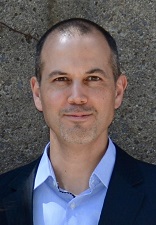}}]{Martin P. Robillard}
is a Professor of Computer Science at McGill University.
His research investigate how to facilitate the discovery and acquisition of technical, design, and domain knowledge to support the development of software systems.
He served as the Program Co-Chair for the 20th ACM SIGSOFT International Symposium on the Foundations of Software Engineering (FSE 2012) and the 39th ACM/IEEE International Conference on Software Engineering (ICSE 2017).
He received his Ph.D. and M.Sc. in Computer Science from the University of British Columbia and a B.Eng. from \'{E}cole Polytechnique de Montr\'{e}al.
\end{IEEEbiography}

\end{document}